\documentclass[9pt]{book}

\usepackage[dvips]{graphicx,color}
\usepackage{makeidx,universe}

\makeauthorindex
\makeindex

\BookTitle{The proceedings of the Physics of Accreting Compact Binaries}

\CopyRight{\copyright 2010 by Universal Academy Press, Inc.}

\begin{document} 

\pagenumbering{arabic}

\chapter{Cyclotron Modeling of the Polar CP Tuc}

\author{\raggedright \baselineskip=10pt%
{\bf Claudia V. Rodrigues,$^{1}$
Joaquim E.~R. Costa,$^{1}$
Karleyne M.~G. Silva,$^{1}$
Cleber A. de Souza,$^{1}$
Deonisio Cieslinski,$^{1}$
and 
Gabriel R. Hickel$^{2}$}\\ 
{\small \it %
(1) Divis\~ao de Astrof\'\i sica, Instituto Nacional de Pesquisas Espaciais, Av. dos Astronautas 1758,  12227-010 S\~ao Jos\'e dos Campos-SP, Brazil \\
(2) Instituto de Ci\^encias Exatas, Universidade Federal de Itajub\'a, Itajub\'a, Brazil
}
}


\AuthorContents{Claudia V. Rodrigues, Joaquim E.~R. Costa, Karleyne M.~G. Silva, Cleber A. de Souza, Deonisio Cieslinski, and Gabriel R. Hickel}

\AuthorIndex{Rodrigues}{C.~V.} 
\AuthorIndex{Silva}{K.~M.~G.}
\AuthorIndex{Costa}{J.~E.~R.} 
\AuthorIndex{de Souza}{C.~A.} 
\AuthorIndex{Cieslinski}{D.} 
\AuthorIndex{Hickel}{G.~R.} 

     \baselineskip=10pt
     \parindent=10pt

\section*{Abstract} 

CP Tuc (AX~J2315-592) is a polar for which two accretion geometries have been put forward. They differ in the location of the accretion column on the white-dwarf surface. In this study, we present new multiband optical photometric and polarimetric data to study the post-shock position. These data are used as input for {\sc cyclops}, our new code for modeling the cyclotron emission from post-shock regions of polars. {\sc cyclops} has a full 3-D treatment of the radiative transfer and is a powerful tool to study the physical properties of the emitting region. We find reasonable fits for the optical data of CP Tuc in \textit{both} scenarios. We suggest that it should be possible to constrain the location of the base of the accreting column by modeling the X-ray light curves.

\section{Introduction} 

Cataclysmic variables (CVs) are compact binaries consisting of a white-dwarf (primary) and a late-type main-sequence star. The secondary star fills its Roche lobe, losing material to the primary by the inner Lagrangian point, ${\rm L_1}$. Due to its angular momentum and viscous processes, this material forms an accretion disk around the white dwarf (WD). Polars, also called AM Her systems after their prototype, are CVs without disks. The disks are inhibited by the strong magnetic field of the primary which ranges from 10 to 200 MG in the WD surface. In these systems, the material from ${\rm L_1}$ follows a ballistic trajectory on the orbital plane (horizontal stream) up to the magnetic coupling region located farther from the circularization radius. From the coupling region on, the gas flow effectively traces the geometry of the magnetic field, forming an accretion column. Another consequence of the strong primary magnetic field is the synchronization of the white-dwarf rotation with the orbital revolution. 

Near the white-dwarf surface, the accreting flux forms a shock front. The accumulated material in the region between the shock and the WD surface is called post-shock region. This region cools down by the cyclotron and \textit{ bremsstrahlung} radiation and dominates the binary emission. A large fraction of the optical flux in polars is cyclotron in origin. Hence, polars are strongly polarized. More than that, the cyclotron emission is anisotropic and consequently the observed flux and polarization depend strongly on orbital phase. This makes polarimetry a fundamental tool to study AM Her systems. 

Polars are also important Galactic X-ray emitters due to their hot post-shock region. As a consequence, high-energy satellites have discovered many polars, contributing significantly to the number of known systems.

The source AX~J2315-592 was, since its discovery by the Japanese satellite ASCA, classified as a polar \cite{cvr_mis95}. Some days after that report, its classification was confirmed by spectroscopy of the optical counterpart, CP Tuc \cite{cvr_tho95}. A deeper analysis of these optical data showed up emission lines with the two components usually found in polars: one narrow, associated with the illuminated secondary, and one broad, originated in the accretion column near the white dwarf  \cite{cvr_tho96}. The zero point of the spectroscopic ephemeris was defined at the blue-to-red crossing of the narrow component, so at the inferior conjunction of the secondary. 

The ASCA data was used to construct light-curves in three X-ray energy intervals and spectra in two ranges of orbital phases for AX~J2315-592 \cite{cvr_mis96}. It was suggested that the modulation of the X-ray light-curves is caused by absorption produced by the accretion column \cite{cvr_mis96}. Indeed, the spectrum could be fitted considering absorption in the fainter phases. The observed X-ray minimum coincides with the inferior conjunction of the secondary. The X-ray data are compatible with those of an intermediate polar (IP) \cite{cvr_mis96}. 

Further X-ray observations and optical photometry and polarimetry were obtained for CP Tuc \cite{cvr_ram99}. These data show no periodicity in addition to the orbital one, what discarded an IP classification. The presence of white light circular polarization, reaching 10\% and modulated with the orbital phase, confirmed the system as a polar. The optical flux and polarization were modeled considering a post-shock region that is self-eclipsed in part of orbital cycle \cite{cvr_ram99}. These authors suggested that the modulation of the X-ray light curve is caused by partial occultation of the emitting region. 

In this study, we present new optical multiband photo-polarimetric data of CP Tuc. These data are used to model the post-shock region of this system. We discuss if it is possible to discriminate between the above two possible geometries with this data set.
 
\section{OBSERVATIONS AND DATA REDUCTION}
\label{sec-obs}

The observations of CP Tuc were performed with the Perkin-Elmer 1.6~m telescope at the {\it Observat\'orio do Pico dos Dias} (OPD) operated by the {\it Laborat\'orio Nacional de Astrof\'\i sica} (LNA), Brazil. We used a CCD camera modified by a polarimetric module \cite{cvr_mag96}. The CCD used was an EEV front-illuminated thick device, with $770\times1152$ pixels. The observations were done in 1997 August 29 to 31 in the R$_{\rm C}$ and I$_{\rm C}$ bands using 90s of exposure time.

\begin{figure}[t]

 \begin{tabular}{cc}
  \begin{minipage}{.47\hsize}
   \begin{center}
     \includegraphics[width=1.05\textwidth]{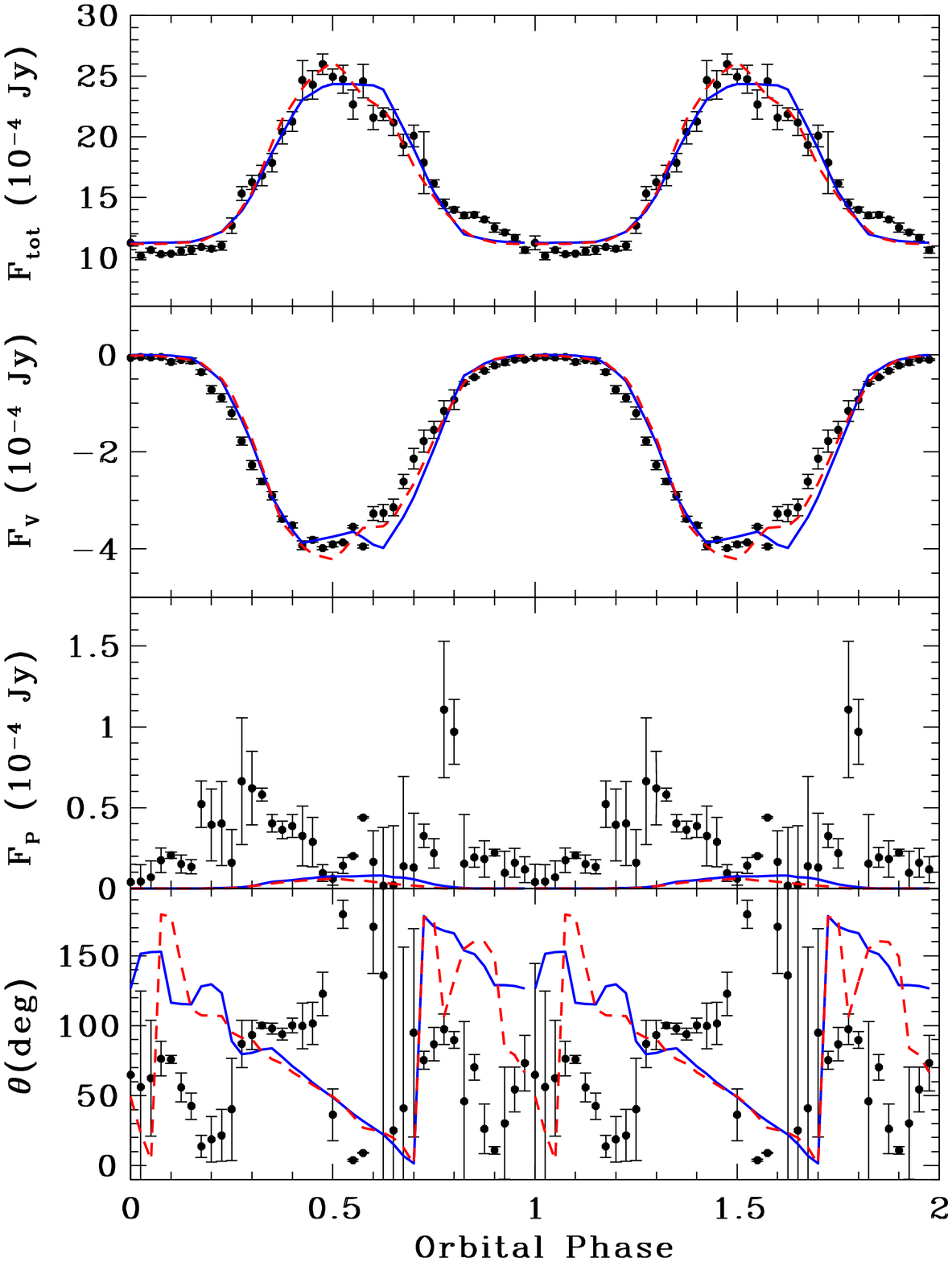}
   \end{center}
  \end{minipage}

  \begin{minipage}{.47\hsize}
   \begin{center}
     \includegraphics[width=1.05\textwidth]{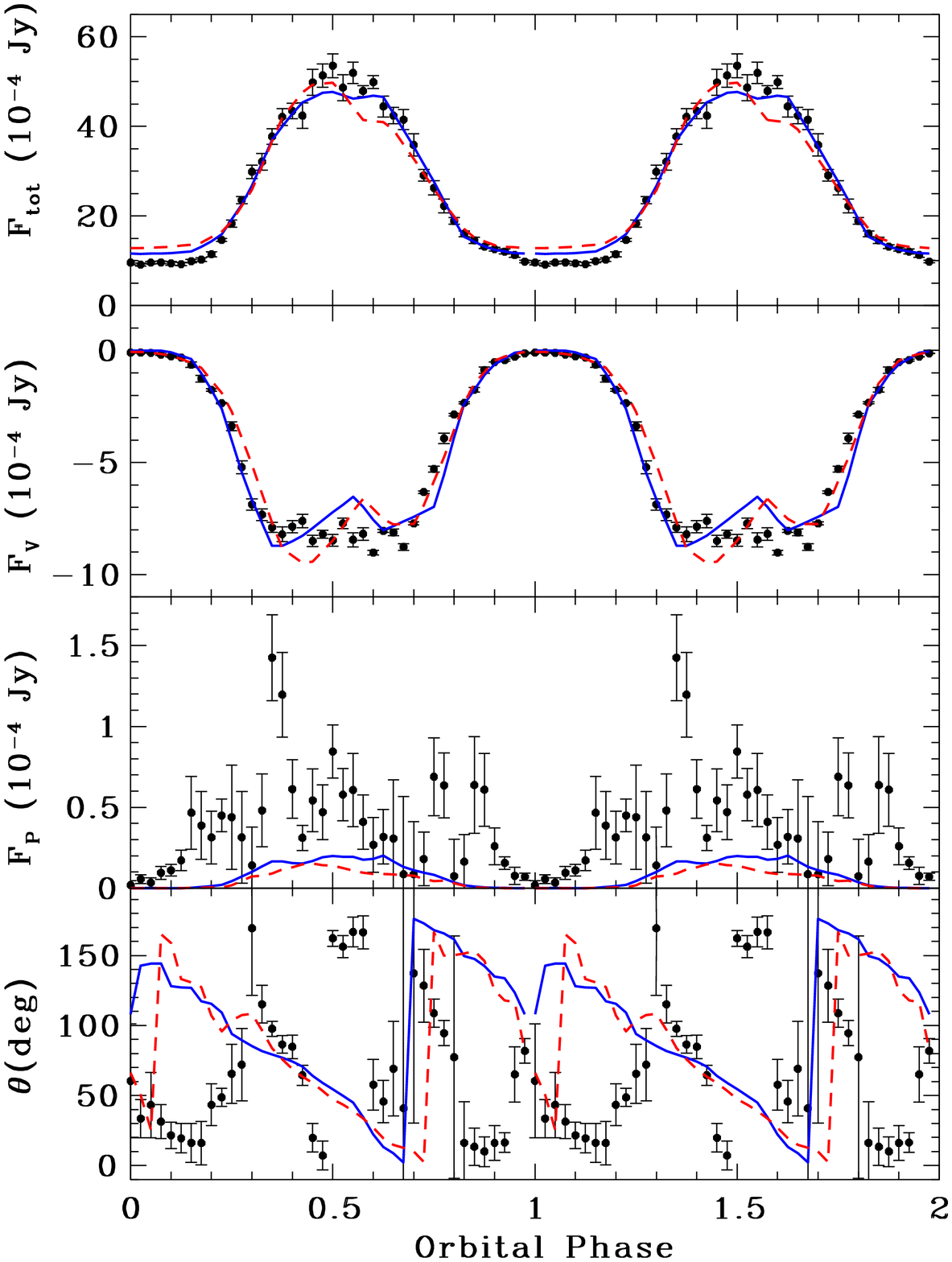}
   \end{center}
  \end{minipage}
 \end{tabular}
 \caption{Data and models of CP Tuc in R$_{\rm C}$ (left) and I$_{\rm C}$ (right) panels. From top to bottom, the panels show the total flux, the circularly polarized flux, the linearly polarized flux and the position angle of the linear polarization. The points correspond to our observations and the lines correspond to the two models presented in Table \ref{cvr_tab}: model 1 (solid blue line) and model 2 (dashed red line). See text for details.}
 \label{cvr_fig_ax23}
\end{figure}

The images were reduced following standard procedures using {\sc iraf}\footnote{{\sc iraf} is distributed by
National Optical Astronomy Observatories, which is operated by the Association of Universities for Research in Astronomy, Inc., under contract with the National Science Foundation.}. The counts obtained from aperture photometry of the ordinary and extraordinary beams were used to calculate the polarization \cite{cvr_mag84,cvr_rod98}. We made use of the package {\sc pccdpack} \cite{cvr_per00}, an set of {\sc iraf} routines specifics to polarimetry.
Each night, we observed linear polarization standard stars to calibrate the system and estimate the instrumental linear polarization. The measured polarization of the unpolarized standard stars were consistent with zero within the errors. Linear polarized and unpolarized standard stars also show insignificant values of circular polarization. Consequently no instrumental correction was applied. Data using a Glan filter were also collected to estimate the instrumental efficiency. They indicate that no correction is needed. Unfortunately, we could not calibrate our measurements in order to know the correct sign of the circular polarization. However, we can distinguish polarization of different signs. We choose the convection that makes the signs consistent with the previous white-light polarimetric measurements \cite{cvr_ram99}.

Photometry was done combining the counts in the two beams. We performed differential photometry of CP Tuc using as reference the star USNO~B1.0~0308$-$0806694 (${\rm R_2}$~=~14.4~mag; ${\rm I}$~=~13.57~mag). The conversion between ${\rm R_2}$ and Landolt's ${\rm R_C}$ for this object indicates a difference of only 0.03 mag \cite{cvr_kid03}. We assume that the USNO $I$ magnitude is the same of that in the Landolt's system.

Figure \ref{cvr_fig_ax23} shows our data of AX~J2315-592, after converting magnitudes and polarization to fluxes. The panels shows from top to bottom: the total flux; the circularly polarized flux; the linearly polarized flux; and the position angle of the linear polarization. The linear polarization was not subtracted from the observational bias. The data were phased using the more recent ephemeris of CP Tuc \cite{cvr_ram99}. 

\section{CYCLOTRON MODELING: RESULTS AND DISCUSSION}

The data presented in the previous section were modeled using {\sc cyclops} \cite{cvr_cr09}. {\sc cyclops} is a three-dimension code to solve the radiative transfer in post-shock regions of polars. Already implemented is the cyclotron emission necessary to model the optical emission of polars as well as the absorption due to bremsstrahlung. The emitting region is divided in small elements of volume and each of them can have different physical properties, namely, magnetic field (intensity and direction), electronic temperature and density. The radiative transfer is solved in steps using appropriated equations for study optically thick regions. The magnetic field of white dwarf is considered as a dipole and the distribution of matter follows the prescription of post-shock regions in polars \cite{cvr_cro99}. To fit observational data, we used the algorithms \textit{amoeba} and \textit{pikaia} \cite{cvr_cha95}. Firstly, we run pikaia with a very broad range of parameters. Then, the best models are refined using the amoeba algorithm.

\begin{table}[ht]

\small

\caption{Parameters of two good models to CP Tuc optical data.} 

\begin{center}

\begin{tabular}{l|cc} \hline

Parameters & Model 1 &  Model 2\\   \hline

inclination, deg         & 29.5    & 22   \\

$\beta$, deg     & 23.8 & 41.6 \\
$\Delta_{long}, deg$ & 18.1 & 12.7 \\
$\Delta_R$ & 0.32 & 0.52 \\
h, $R_{WD}$ & 0.10 & 0.22 \\
$B_{pole}$, MG & 6 & 6.1 \\
$B_{lat}$, deg & 85 & 35 \\
$B_{long}$, deg & 41 & 336 \\
$T_{max}$, keV & 66 & 102 \\
$N_{max}$, cm-3 (log) & 15.7 & 15 \\
\hline
\end{tabular}

\label{cvr_tab}
\end{center}

A concise explanation of the {\sc cyclops} parameters is as follows: $\beta$ is the angle between the rotation axis and the center of the northern region; $\Delta_{long}$ and $\Delta_R$ define the size of the coupling region and, hence, the footprint size of the emitting region in the WD surface; h is the height of emitting region in units of  WD radius; $B_{pole}$ is the intensity of the magnetic field on the WD pole; $B_{lat}$ and $B_{long}$ define the direction of the dipole axis; $T_{max}$ and $N_{max}$ are the maximum values of the electronic temperature and density, respectively. Please, see the original publication \cite{cvr_cr09} for details.
\end{table}

\begin{figure}[ht]
   \begin{center}
     \includegraphics[width=1.05\textwidth]{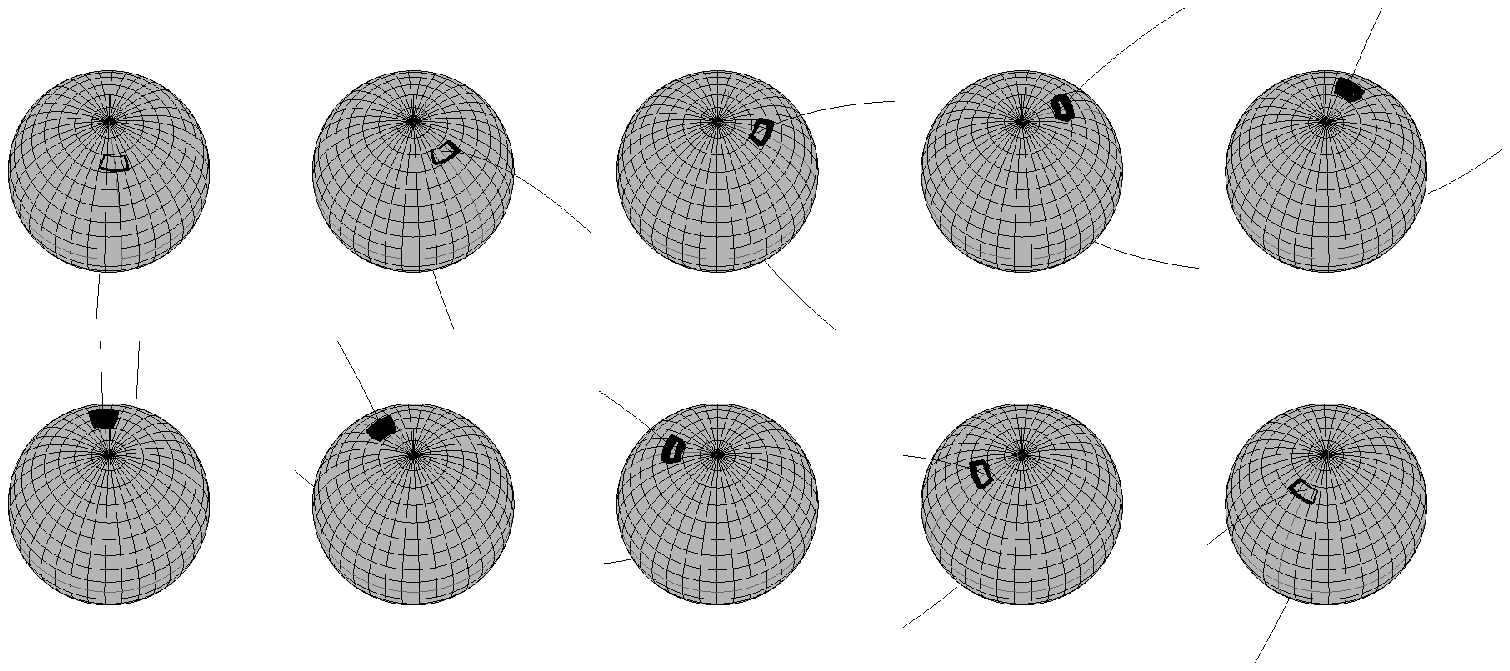}
      \caption{View of the emitting region in Model 1 from orbital phases 0. to 0.9 in 0.1 steps. Only the ``walls`` are shown. The curved line represents the magnetic field line that intercepts the center of the emitting region. See Table \ref{cvr_tab} for the models parameters.}
      \label{cvr_fig_m1}
   \end{center}

   \begin{center}
     \includegraphics[width=0.95\textwidth]{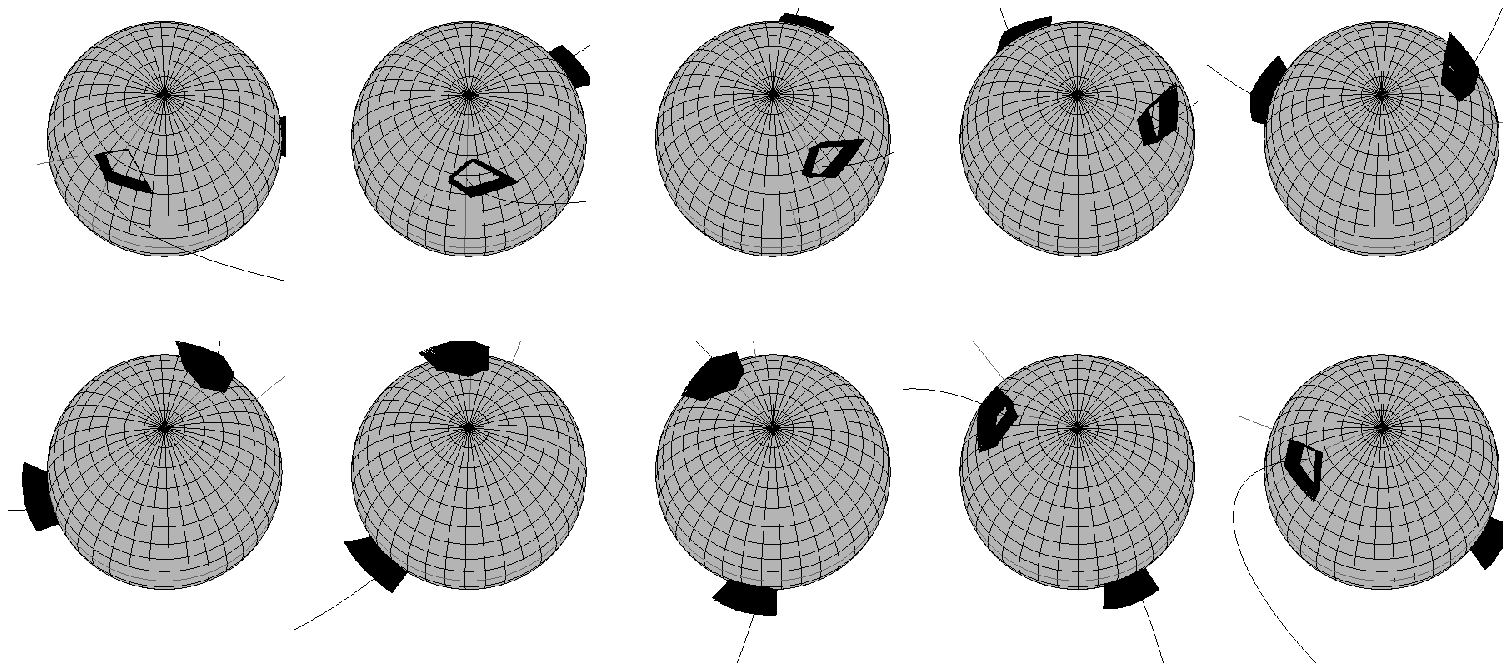}
 \caption{The same as Figure \ref{cvr_fig_m1} for Model 2. The emitting region is the southern region (right on the first panel). No emission is considered from the region located at the northern hemisphere.}
\label{cvr_fig_m2}
   \end{center}
\end{figure}

In searching for a good model, we obtain naturally two classes of models. An example of each class is shown in Figure \ref{cvr_fig_ax23}. The parameters of each model are presented in Table \ref{cvr_tab}. Views of the emitting region as a function of the orbital phase are presented in Figures \ref{cvr_fig_m1} and \ref{cvr_fig_m2} for each model.
Both models represent equally well the data. We did not search for a proper fit of the linear polarization, since the linear polarization is small and not correctly unbiased. Even so, both models show small values of the linear polarization, consistently with the observational data. Model 1 represents a geometry in which a post-shock region is always visible (Figure \ref{cvr_fig_m1}). The region corresponding to the other footprint of the magnetic lines is always eclipsed by the white-dwarf. In Model 2, on the other hand, the two footprint of the magnetic lines can be seen (Figure \ref{cvr_fig_m2}). In our modeling, only the southern region is emitting. This region is partially self-eclipsed in some orbital phases.

Interestingly, each model represents one of the scenarios proposed to explain the X-ray light curves of CP Tuc. Model 1 is consistent with X-ray light curves modulated by absorption produced in the upper accreting column \cite{cvr_mis96}. In this model, the pre-shock region intercepts the line-of-sight in a large range of orbital phases centered at zero phase (see Figure \ref{cvr_fig_m1}). {\sc cyclops} takes into account this absorption. However, it is not important at optical wavelengths in Model 1. Model 2 represents the scenario in which the X-ray modulation is caused by self-eclipse of the post-shock region \cite{cvr_ram99}.

What is the geometry that best represents the accretion column of CP Tuc? A possible solution is to use the X-ray light curves. Until now, no modeling of the X-ray light curves of CP Tuc has been attempted in the literature: only qualitative reasonings have been presented.

We are currently working on the inclusion of X-ray emission at {\sc cyclops} (please, see Silva et al., this proceedings). {\sc cyclops-X } includes the absorption of the upper column considering consistently the geometry defined by the magnetic lines and the quantity of matter in the pre-shock region.

\section{CONCLUSIONS}

We present new optical polarimetric and photometric observations of the polar CP Tuc in two bands: R$_{\rm C}$ and I$_{\rm C}$. The {\sc cyclops} code is used to model these data and study the physical and geometrical properties of the post-shock region of CP Tuc. We find good models in the two geometrical scenarios previously proposed for the post-shock region: (1) an region always visible; (2) a region partially self-eclipsed in some orbital phases.  We suggest that it is possible to settle this question checking which model is able to simultaneously reproduce the X-ray and optical data. We will do that using {\sc cyclops-X}, a new tool to study the optical and X-ray emission of polars.

\section*{ACKNOWLEDGEMENTS}

This work was partially supported by Fapesp (CVR: Proc. 2010/06096-1; KMGS: 2008/09619-5). 



\end{document}